# I/O-Efficient Data Structures for Colored Range and Prefix Reporting

Kasper Green Larsen

MADALGO*

Aarhus University, Denmark

larsen@cs.au.dk

Rasmus Pagh

IT University of Copenhagen

Copenhagen, Denmark

pagh@itu.dk

**Abstract**

Motivated by information retrieval applications, we consider the *one-dimensional colored range reporting* problem in rank space. The goal is to build a static data structure for sets $C_1, \ldots, C_m \subseteq \{1, \ldots, \sigma\}$ that supports queries of the kind: Given indices $a, b$, report the set $\bigcup_{a \leq i \leq b} C_i$.

We study the problem in the I/O model, and show that there exists an optimal linear-space data structure that answers queries in $O(1 + k/B)$ I/Os, where $k$ denotes the output size and $B$ the disk block size in words. In fact, we obtain the same bound for the harder problem of *three-sided orthogonal range reporting*. In this problem, we are to preprocess a set of $n$ two-dimensional points in rank space, such that all points inside a query rectangle of the form $[x_1, x_2] \times (-\infty, y]$ can be reported. The best previous bounds for this problem is either $O(n \lg_B^2 n)$ space and $O(1 + k/B)$ query I/Os, or $O(n)$ space and $O(\lg_B^{(h)} n + k/B)$ query I/Os, where $\lg_B^{(h)} n$ is the base $B$ logarithm iterated $h$ times, for any constant integer $h$. The previous bounds are both achieved under the *indivisibility assumption*, while our solution exploits the full capabilities of the underlying machine. Breaking the indivisibility assumption thus provides us with cleaner and optimal bounds.

Our results also imply an optimal solution to the following *colored prefix reporting* problem. Given a set $S$ of strings, each $O(1)$ disk blocks in length, and a function $c : S \to 2^{\{1,\ldots,\sigma\}}$, support queries of the kind: Given a string $p$, report the set $\bigcup_{x \in S \cap p*} c(x)$, where $p*$ denotes the set of strings with prefix $p$. Finally, we consider the possibility of top-$k$ extensions of this result, and present a simple solution in a model that allows non-blocked I/O.

# 1 Introduction

A basic problem in information retrieval is to support prefix predicates, such as `datab*`, that match all documents containing a string with a given prefix. Queries involving such a predicate are often resolved by computing a list of all documents satisfying it, and merging this list with similar lists for other predicates (e.g. inverted indexes). Recent overviews can be found in e.g. [29, 12]. To our best knowledge, existing solutions either require super-linear space (e.g. storing all answers) or report a multi-set, meaning that the same document may be reported many times if it has many words with the given prefix. In range reporting terminology we are interested in the *colored* reporting problem, where each color (document) may match several times, but we are only interested in reporting each color once.

A related problem is that of *query relaxation*. When no answers is produced on a query for some given terms, an information retrieval system may try to "relax" some of the search conditions to produce near-matches. For example, a search for "colour television" may be relaxed such that also documents containing "color" and/or "TV" are regarded as matches, or further relaxed to also match "3D" and "screen". In general, terms may be arranged as leaves in a tree (possibly with duplicates), such that inner nodes correspond to a natural relaxation of the terms below it. When answering a query we are interested in the documents containing a term in some subtree. Again, this reduces to a colored 1D range reporting problem.

## 1.1 Model of Computation

In this paper we study the above problems in the I/O-model [6] of computation. In this model, the input to a data structure problem is assumed too large to fit in main memory of the machine, and thus the data structure must reside on disk. The disk is assumed infinite, and is divided into disk blocks, each consisting of $B$ words of $\Theta(\lg n)$ bits each. We use $b = \Theta(B \lg n)$ to denote the number of bits in a disk block.

To answer a query, the query algorithm repeatedly reads disk blocks into the main memory (an I/O) of size $M$ words, and based on the contents of the read blocks it finally reports the answer to the query. The cost of answering a query is measured in the number of I/Os performed by the query algorithm.

**The indivisibility assumption.** A common assumption made when designing data structures in the I/O-model is that of *indivisibility*. In this setting, a data structure must treat input elements as atomic entities, and has no real understanding of the bits constituting the words of a disk block. There is one main motivating factor for using this restriction: It makes proving lower bounds significantly easier. This alone is not a strong argument why upper bounds should be achieved under the indivisibility assumption, but it has long since been conjectured that for most natural problems, one cannot do (much) better without the indivisibility assumption. In this paper, we design data structures without this assumption, which allows us to achieve optimal bounds for some of the most fundamental range searching problems, including the problems above.

## 1.2 Our contributions

In this paper we present the first data structures for colored prefix reporting queries that simultaneously have linear space usage and optimal query complexity in the I/O model [6]. More precisely, our data structure stores sets $C_1, \ldots, C_m \subseteq \{1, \ldots, \sigma\}$ and supports queries of the kind: Given indices $a, b$, report the set $\bigcup_{a \leq i \leq b} C_i$. If the reported set has size $k$, then the number of I/Os used to answer the query is $O(1 + k/B)$. If we let $n = \sum_{i=1}^{m} |C_i|$ denote the total data size, then the space usage of the data structure is $O(n)$ words, i.e. linear in the size of the data.

In fact, in Section 2 we present an optimal solution to the harder and very well-studied range searching problem of *three-sided orthogonal range reporting* in two-dimensional rank-space, and then use a known reduction [19] to get the above result (Section 3). Given a set $S$ of $n$ points from the grid $[n] \times [n] = \{1,\ldots,n\} \times \{1,\ldots,n\}$, this problem asks to construct a data structure that is able to report all points inside a query rectangle of the form $[x_1, x_2] \times (-\infty, y]$. We consider the static version of the problem, where updates to the set $S$ are not required. We note that optimal solutions to this problem have been known for more than a decade on the word-RAM, but these solutions are all inherently based on random accesses. The query time of these data structures thus remains $O(1+k)$ when analysed in the I/O-model, which falls short of the desired $O(1 + k/B)$ query cost.

One of the key ideas in obtaining our results for three-sided range reporting is an elegant combination of Fusion Trees and tabulation, allowing us to make a dynamic data structure partially persistent, *free of charge*, when the number of updates to the data structure is bounded by $b^{O(1)}$. We believe several of the ideas we develop, or variations thereof, may prove useful for many other natural problems.

Finally, in Section 4 we consider the top-$k$ variant, where we only need to report the first $k$ colors in the set, and $k$ is a parameter of the data structure. In a *scatter I/O* model that allows $B$ parallel word accesses we get a data structure with optimal query complexity also for this problem. This data structure uses space $O(n + mB)$, which is $O(n)$ if the average size of $C_1, \ldots, C_m$ is at least $B$. Since the scatter I/O model abstracts the parallel data access capabilities of modern hardware, we believe that it is worth investigating further, and our results here may be a first step in this direction.

**Notation.** We always use $n$ to denote the size of data in number of elements. For colored prefix reporting this means that we have $m$ subsets of $\{1, \ldots, \sigma\}$ of total size $n$. As mentioned, we use $b$ to denote the number of bits in a disk block. Since a block may store $B$ pointers and input elements, we make the natural assumption that $b = \Theta(B \lg n)$, i.e., each disk block consists of $B$ words, where a word is $\Theta(\lg n)$ bits.

## 1.3 Related work

The importance of three-sided range reporting is mirrored in the number of publications on the problem, see e.g. [21, 17] for the pointer machine, [5, 9, 11, 3, 4] for the cache-oblivious and [8, 26, 14] for the word-RAM model. One of the main reason why the problem has seen so much attention stems from the fact that range searching with more than three sides no longer admits linear space data structures with polylogarithmic query cost and a linear term in the output size. Thus three-sided range searching is the "hardest" range searching problem that can be solved efficiently both in terms of space and query cost, and at the same time, this has proved to be a highly non-trivial task.

The best I/O model solution to the three-sided range reporting problem in 2-d, where coordinates can only be compared, is due to Arge et al. [10]. Their data structure uses linear space and answers queries in $O(\lg_B n + k/B)$ I/Os. This is optimal when coordinates can only be compared. Nekrich reinvestigated the problem in a setting where the points lie on an integer grid of size $U \times U$. He demonstrated that for such inputs, it is possible to achieve $O(\lg_2 \lg_B U + k/B)$ query cost while maintaining linear space [23]. This bound is optimal by a reduction from predecessor search. His solution relies on indirect addressing and arithmetic operations such as multiplication and division. In the same paper, he also gave several data structures for the case of input points in rank space. The first uses linear space and answers queries in $O(\lg_B^{(h)} n + k/B)$ I/Os, where $\lg_B^{(h)} n$ is the base $B$ logarithm iterated $h$ times, for any constant integer $h$. The second uses $O(n \lg_B^* n)$ space and answers queries in $O(\lg_B^* n + k/B)$ I/Os, and the final data structure uses $O(n \lg_B^2 n)$ space and answers queries in optimal $O(1 + k/B)$ I/Os. All these data structures use only comparisons and indirect addressing.

Higher-dimensional orthogonal range reporting has also received much attention in the I/O model, see e.g. [27, 1, 2, 24]. The best current data structures for orthogonal range reporting in $d$-dimensional space ($d \geq 3$), where coordinates can only be compared, either answers queries in $O(\lg_B n (\lg n / \lg \lg_B n)^{d-2} + k/B)$ I/Os and uses $O(n(\lg n / \lg \lg_B n)^{d-1})$ space, or answers queries in $O(\lg_B n \lg^{d-3} n + k/B)$ I/Os and uses $O(n(\lg n / \lg \lg_B n)^3 \lg^{d-3} n)$ space [2].

All these results in some sense do not exploit the full power of the underlying machine. While this provides for easier lower bounds, it should be clear when comparing to our results, that this approach might come at a cost of efficiency. Finally, we note that recent work by Iacono and Pătrașcu [20] also focuses on obtaining stronger upper bounds (for dynamic dictionaries) in the I/O model by abandoning the indivisiblity assumption.

## 2 Three-Sided Orthogonal Range Reporting

In this section we describe our data structure for three-sided orthogonal range reporting. Recall that in this problem we are interested in reporting the points in $[x_1, x_2] \times (-\infty, y]$. If there are $k$ such points, our data structure answers queries in $O(1 + k/B)$ I/Os and uses linear space for input points in rank space, i.e., when the input points lie on the grid $[n] \times [n]$. We set out with a brief preliminaries section, and then describe our data structure in Section 2.2.

### 2.1 Preliminaries

In this section, we briefly discuss two fundamental data structures that we make use of in our solutions, the Fusion Tree of Fredman and Willard [16] and the External Memory Priority Search Tree (EM-PST) of Arge et al. [10]. For the EM-PST, we present a parametrized version of the original data structure, while for the Fusion Tree, we merely state the result and use it as a black box.

**Fusion Tree.** Allowing for full access to the bits in a disk block yields more efficient data structure solutions to several fundamental problems. In this paper, we use the Fusion Tree of Fredman and Willard[1]:

**Lemma 1** *There exists a linear space data structure that supports predecessor search in $O(\lg_b n)$ I/Os on $n$ elements in a universe of size $u$, where $\lg u < b$.*

The requirement $\lg u < b$ ensures that we can store an element in a single disk block. In the original Fusion Tree data structure, much care is taken to implement various operations on words in constant time. However this is trivialized in the I/O model, since we can do arbitrary computations on a disk block in main memory free of charge.

**External Memory Priority Search Tree.** The EM-PST is an external memory data structure that answers three-sided range queries. We describe the basic layout of the EM-PST, parametrized with a *leaf*-parameter $\ell$ and a *branching*-parameter $f$.

An EM-PST is constructed from an input set of $n$ points in the following recursive manner: Select the $B$ points with smallest $y$-coordinate among the remaining points and place them in a root node. Then partition the remaining points into $f$ equal-sized consecutive groups wrt. the $x$-coordinates of the points.

---
[1] Strictly speaking, we use what Fredman and Willard refer to as $q$-heaps, which generalize the more well-known fusion trees.

Recurse on each group, and let the recursively constructed trees be the children of the root node, ordered by $x$-coordinates ($f - 1$ splitter keys are stored in a Fusion Tree at the root). We end the recursion when the number of points in a subproblem drops to between $\ell$ and $f\ell + B$, and instead create a leaf node. Note that the EM-PST is both a heap on the $y$-coordinates and a search tree on the $x$-coordinates, hence it's name. Furthermore, the height of the tree is $O(\lg_f(n/\ell))$, and there are $O(n/\ell)$ nodes in the tree.

The original data structure of Arge et al. has leaf-parameter $\ell = B$ and branching parameter $f = B$. They augment each node of the tree with a number of auxiliary data structures to support fast queries. We omit the description of these data structures, as we will only use the basic layout of the tree in our solutions.

## 2.2 Data Structure

In this section, we describe our optimal data structure for three-sided range reporting in rank space. At a high level, our data structure places the points in an EM-PST, and augments each leaf with auxiliary data structures. These allow efficient reporting of all points in a query range $[x_1, x_2] \times (-\infty, y]$ that are associated with nodes on the path from the root to the corresponding leaf. Since the number of points associated with nodes on such a path is rather small, we are able to answer these queries in $O(1 + k/B)$ I/Os. To report the remaining points in a query range, we exploit that for each node not on the path to either of the two leaves containing the predecessors of $x_1$ and $x_2$, either all the points associated with the node have an $x$-coordinate in the query range, or none of them have. We now give the full details of our solution.

**The Data Structure.** Suppose there exists a base data structure for three-sided range reporting that uses linear space and answers queries in $O(1 + k/B)$ I/Os when constructed on $b^{O(1)}$ points with coordinates on the grid $[n] \times [n]$. Let $S$ be the input set of $n$ points, and construct on $S$ the EM-PST layout described in Section 2.1, using branching parameter $f = 2$ and leaf-parameter $\ell = B \lg^2 n$. Denote the resulting tree by $T$. For each leaf $v$ in $T$ we store two auxiliary data structures:

1. A base data structure on the $O(f\ell + B) = O(B \lg^2 n) = O(b^2)$ points associated with $v$.

2. For each ancestor $w$ of $v$, we store a base data structure on the $O(fB \lg n) = O(b)$ points associated with all nodes that are either on the path from $w$ to the parent of $v$, or is a sibling of a node on the path. We furthermore augment each of the points with the node it comes from.

Finally, we augment $T$ with one additional auxiliary data structure. This data structure is a simple array with $n$ entries, one for each $x$-coordinate. The $i$th entry of the array stores a pointer to the leaf containing the $x$-predecessor of $i$ amongst points associated with leaves of $T$. If $i$ has no such predecessor, then we store a pointer to the first leaf of $T$.

**Query Algorithm.** To answer a query $[x_1, x_2] \times (-\infty, y]$, we first use the array to locate the leaf $u_1$ containing the predecessor of $x_1$, and the leaf $u_2$ containing the predecessor of $x_2$. We then locate the lowest common ancestor $\text{LCA}(u_1, u_2)$ of $u_1$ and $u_2$. Since $T$ is a balanced binary tree, $\text{LCA}(u_1, u_2)$ can be found with $O(1)$ I/Os if each node contains $O(\lg n)$ bits describing the path from the root.

We now query the base data structures stored on the points in $u_1, u_2$ and their sibling leaves. This reports all points in the query range associated with those nodes. We then query the second base data structure stored in $u_1$, setting $w$ to the root of $T$ (see 2. above), and the second base data structure stored in $u_2$ setting $w$ to that grandchild of $\text{LCA}(u_1, u_2)$ which is on the path to $u_2$. Observe that this reports all points in the query range that are either associated to nodes on the paths from the root to $u_1$ or $u_2$, or to nodes that have a sibling on the paths.

We now exploit the heap and search tree structure of $T$: For a node $v$ not on the paths to $u_1$ and $u_2$, but with a sibling on one of them, we know that either the entire subtree rooted at $v$ stores only points with an $x$-coordinate inside the query range, or none of the points in the subtree are inside the query range. Furthermore, the heap ordering ensures that the points with smallest $y$-coordinate in the subtree are associated to $v$. Thus if not all $B$ points associated to $v$ was reported above, then no further points in the subtree can be inside the query range. We thus proceed by scanning all the reported points above, and for each node $v$ not on the paths to $u_1$ and $u_2$, we verify whether all $B$ associated points were reported. If this is the case, we visit the subtree rooted at $v$ in the following recursive manner:

If the children of $v$ are not leaves, we scan the $B$ points associated to both of them, and report those with a $y$-coordinate inside the query range. If all $B$ points associated to a child are reported, we recurse on that child. If the children are leaves, we instead query the base data structure stored on the associated points and terminate thereafter.

As a side remark, observe that if we mark the point with largest $y$-coordinate in each node, then all $B$ points associated to a node with a sibling on one of the two paths are reported, iff the marked point associated to the node is reported. Thus the above verification step can be performed efficiently.

**Analysis.** The base data structures stored on the points in the leaves uses $O(n)$ space in total, since each input point is stored in at most one such data structure. There are $O(n/\ell) = O(n/(B \lg^2 n))$ leaves, each storing $O(\lg_f n) = O(\lg n)$ data structures of the second type. Since each such data structure uses $O(B \lg n)$ space, we conclude that our data structure uses linear space in total.

The query cost is $O(1)$ I/Os for finding $u_1, u_2$ and $\text{LCA}(u_1, u_2)$. Reporting the points in $u_1, u_2$ and their siblings costs $O(1 + k/B)$ I/Os. Querying the second base data structures in $u_1$ and $u_2$ also costs $O(1 + k/B)$ I/Os. Finally observe that we only visit a node not on the paths to $u_1$ or $u_2$ if all $B$ points associated to it are reported. Since we spend $O(f) = O(1)$ I/Os visiting the children of such a node, we may charge this to the output size, and we conclude that our data structure answers queries in $O(1 + k/B)$ I/Os, assuming that our base data structure is available.

**Lemma 2** *If there exists a linear space data structure for three-sided range reporting that answers queries in $O(1 + k/B)$ I/Os on $b^{O(1)}$ points on the grid $[n] \times [n]$, then there exists a linear space data structure for three-sided range reporting on $n$ points in rank space, that answers queries in $O(1 + k/B)$ I/Os.*

## 2.3 Base Data Structure

In the following we describe our linear data structure that answers three-sided range queries in $O(1 + k/B)$ I/Os on $b^{O(1)}$ points on the grid $[n] \times [n]$. We use two different approaches depending on the disk block size:

If $B = \Omega(\lg^{1/16} n)$, then we use the EM-PST of Arge et al. This data structure uses linear space and answers queries in $O(\lg_B b^{O(1)} + k/B) = O(1 + k/B)$ I/Os since $b = \Theta(B \lg n) = O(B^{17})$.

The hard case is thus when $B = o(\lg^{1/16} n)$, which implies $b = o(\lg^{17/16} n)$. We first show how we obtain the desired query bound when the number of input points is $O(b^{1/8})$, and then extend our solution to any constant exponent.

### 2.3.1 Very Few Input Points.

In the following we let $m = O(b^{1/8}) = o(\lg^{17/108} n)$ denote the number of input points. Note that if we reduce the $m$ points to rank space, then we can afford to store a table with an entry for every possible input and query pair. Unfortunately, we cannot simply store all answers to queries, as we would need to map the reported points back from rank space to their original coordinates, potentially costing $O(1)$ I/Os per reported point, rather than $O(1/B)$ I/Os. Our solution combines tabulation with the notion of *partial persistence* to achieve the desired query cost of $O(1 + k/B)$ I/Os.

A dynamic data structure is said to be partially persistent if it supports querying any past version of the data structure. More formally, assume that the updates to a dynamic data structure are assigned increasing integer IDs from a universe $[u]$. Then the data structure is partially persistent if it supports, given a query $q$ and an integer $i \in [u]$, answering $q$ as if only updates with ID at most $i$ had been performed. We think of these IDs as time, and say that an update with ID $i$ happened at time $i$.

We observe that a partially persistent insertion-only data structure for 1d range reporting solves the three-sided range reporting problem: Sweep the input points to the three-sided problem from smallest $y$-coordinate to largest. When a point $p = (x, y)$ is encountered, we insert $p$ into the 1d data structure using $x$ as its coordinate, and $y$ as the time of the insertion. To answer a three-sided query $[x_1, x_2] \times (-\infty, y]$, we query the 1d data structure with range $[x_1, x_2]$ at time $y$. This clearly reports the desired points. In the following we therefore devise an insertion-only 1d range reporting data structure, and then show how to make it partially persistent.

**1d Insertion-Only Range Reporting.** Our 1d data structure consists of a simple linked list of disk blocks. Each block stores a set of at most $B - 1$ inserted points, and we maintain the invariant that all coordinates of points inside a block are smaller than all coordinates of points inside the successor block. Note however that we do not require the points to be sorted within each block. Initially we have one empty block.

When a point $p$ is inserted with coordinate $x$, we scan through the linked list to find the block containing the predecessor of $x$. We add $p$ to that block. If the block now contains $B$ points, we split it into two blocks, each containing $B/2$ points. We partition the points around the median coordinate, such that all points inside the first block have smaller coordinates than those in the second. This clearly maintains our invariant.

To answer a query $[x_1, x_2]$ we assume for now that we are given a pointer to the block containing the predecessor of $x_1$. From this block we scan forward in the linked list, until a block is encountered where some point has coordinate larger than $x_2$. In the scanned blocks, we report all points that are inside the query range. Because of the ordering of points among blocks, this answers the query in $O(1 + k/B)$ I/Os.

**Partial Persistence.** We now modify the insertion algorithm to make the data structure partially persistent:

1. When a block splits, we do not throw away the old block, but instead maintain it on disk. We also assign increasing integer IDs to each created block, starting from 1 and incrementing by one for each new block.

2. We store an array with one entry for each block ID. Entry $i$ stores a pointer to the block with ID $i$.

3. Whenever a successor pointer changes, we do not throw away the old value. Instead, each block stores (a pointer to) a Fusion Tree that maintains all successor pointers the block has had. Thus when a successor pointer changes, we insert the new pointer into the Fusion Tree with key equal to the time of the change.

4. Finally, we augment each point with its insertion time.

Note that once a block splits, the old block will not receive further updates. The blocks that have not yet split after a sequence of updates constitutes the original data structure after the same updates, and these are the only blocks we update during future insertions.

We answer a query $[x_1, x_2]$ at time $y$ in the following way: Assume for now that we can find the ID of the block that contained the predecessor of $x_1$ at time $y$. Note that this block might have split at later updates, but is still stored in our data structure. We use the array and the ID to retrieve this block, and from there we simulate the original query algorithm. When the original query algorithm requests a successor pointer, we find the predecessor of $y$ in that block's Fusion Tree, yielding the successor pointer of the block at time $y$. We stop when a block containing a point inserted at time at most $y$ and having coordinate larger than $x_2$ is encountered. For all the blocks scanned through this procedure, we report the points with coordinate in the query range, that was inserted at time at most $y$. This answers the query correctly as the blocks scanned contain all points that would have been scanned if the query was asked at time $y$, except possibly some more points inserted after time $y$.

By Lemma 1 we get that each predecessor search costs $O(1)$ I/Os since we have at most $m = O(b^{1/8})$ successor pointer updates. Thus the total query cost is $O(1+k/B)$ I/Os since the successor of a block at time $y$ contains at least $B/2$ points that were inserted at time at most $y$, which allows us to charge the predecessor search to the output size. To argue that the space of our data structure is linear, we charge splitting a block to the at least $B/2$ insertions into the block since it was created. Splitting a block increases the number of used disk blocks by $O(1)$, so we get linear space usage.

**Three-Sided Range Reporting.** The only thing preventing us from using the partially persistent data structure above to solve the three-sided range reporting problem, is a way to find the ID of the block containing the predecessor of $x_1$ at time $y$. For this, we use tabulation.

The key observation we use for retrieving this ID is that if we reduce the coordinates of the input points for the three-sided range reporting problem to rank space before inserting them into the partially persistent 1d data structure, then only the coordinates of points and the update time of each successor pointer changes, i.e., the ID of the block containing a particular point remains the same. Thus we create a table with an entry for every input and query pair to the three-sided range reporting problem in rank space. Each entry of the table contains the ID of the block in the partially persistent 1d structure from which the query algorithm must start scanning.

To summarize, we solve the three-sided range reporting problem by building the partially persistent 1d data structure described above. We then store two Fusion Trees, one on the $x$-coordinates and one on the $y$-coordinates of the input points. These allows us to reduce a query to rank space. Furthermore, we store a table mapping input and query pairs in rank space to block IDs. Finally, we store an integer that uniquely describes the input point set in rank space (such an integer easily fits in $\lg n$ bits). To answer a query $[x_1, x_2] \times (-\infty, y]$, we use the Fusion Trees to map the query to rank space wrt. the input point set. We then use the unique integer describing the input set, and the query coordinates in rank space, to perform a lookup in our table, yielding the ID of the block containing the predecessor of $x_1$ at time $y$. Finally, we execute the query algorithm for the original query (not in rank space) on the partially persistent 1d data structure, using the ID to locate the starting block.

Since there are $m = O(b^{1/8})$ input points, we get reduction to rank space in $O(1)$ I/Os by Lemma 1. Querying the 1d data structure costs $O(1 + k/B)$ I/Os as argued above, thus we conclude:

**Lemma 3** *There exists a linear space data structure for three-sided range queries on $O(b^{1/8})$ points on the grid $[n] \times [n]$, that answers queries in $O(1 + k/B)$ I/Os.*

### 2.3.2 Few Points

In this section we extend the result of the previous section to give a linear space data structure that answers three-sided range queries in $O(1 + k/B)$ I/Os on $b^{O(1)}$ points when $B = o(\lg^{1/16} n)$.

Let $m = b^{O(1)}$ denote the number of input points. We construct the EM-PST layout described in Section 2.1 on the input points, using branching parameter $f = b^{1/16}$ and leaf parameter $\ell = B$. The height of the constructed tree $T$ is $O(1)$. In each internal node $v$ of $T$, we store the data structure of Lemma 3 on the $O(fB) = O(b^{1/8})$ points associated with the children of $v$.

We answer a query $[x_1, x_2] \times (-\infty, y]$ using an approach similar to the one employed in Section 2.2. We first find the two leaves $u_1$ and $u_2$ containing the predecessors of $x_1$ and $x_2$ among points stored in leaves. This can easily be done using the Fusion Trees stored on the split values in each node of $T$. To report the points in the query range that are associated with a node on the two paths from the root to $u_1$ and $u_2$, we simply scan the points associated with each node on the paths. What remains are the subtrees *hanging off* the query paths. These are the subtrees with a root node that is not on the query paths, but which has a parent on the query paths. These subtrees are handled by first traversing the two paths from $\text{LCA}(u_1, u_2)$ to $u_1$ and $u_2$ (subtrees hanging off at a higher node cannot contain points inside the $x$-range of the query). In each node $v$ on these paths, we query the data structure of Lemma 3 with a slightly modified query range: In a node $v$ on the path to $u_1$, we increase $x_1$ to not include $x$-coordinates of points in the child subtree containing $u_1$. Similarly, on the path to $u_2$, we decrease $x_2$ to not include $x$-coordinates of points in the child subtree containing $u_2$. This can easily be done by a predecessor search on the split values. We finish by recursing into each child from which all $B$ points were reported. Here we again query the data structure of Lemma 3 and recurse on children where all points are reported.

For the subtrees hanging off the query path, we only recurse into a node $v$ if all $B$ points associated to $v$ are reported. There we spend $O(1 + k'/B)$ I/Os querying the data structure of Lemma 3 where $k'$ denotes the output size of the query among points associated to children of $v$. We charge all this to the output size. Finally, the height of $T$ is $O(1)$, and we spend $O(1)$ I/Os in each node on the paths from the root to $u_1$ and $u_2$, thus we conclude that our data structure answers queries in $O(1 + k/B)$ I/Os. For the space, simply observe that each points is stored in only one data structure of Lemma 3. We therefore get

**Lemma 4** *There exists a linear space data structure for three-sided range reporting on $b^{O(1)}$ points on the grid $[n] \times [n]$, that answers queries in $O(1 + k/B)$ I/Os.*

If we combine Lemma 2 and 4 we finally get our main result:

**Theorem 1** *There exists a linear space data structure for three-sided range reporting on $n$ points in rank space, that answers queries in $O(1 + k/B)$ I/Os.*

## 3 Colored range and prefix reporting

**Colored range reporting.** Our optimal solution for three-sided range reporting in rank space immediately gives an optimal solution to the one-dimensional colored range reporting problem in rank space. In this problem, we are given sets $C_1, \ldots, C_m \subseteq \{1, \ldots, \sigma\}$, and are to preprocess them into a data structure that supports queries of the form: Given indices $(a, b)$, report the set $\bigcup_{a \leq i \leq b} C_i$. We think of each $C_i$ as an ordered set of colors, hence the name colored range reporting.

Our optimal solution to this problem follows from a simple and elegant reduction described in [19]. Below, we present the reduction as it applies to our problem.

First think of each set $C_i$ as a set of one-dimensional colored points $\{(i, c) \mid c \in C_i\}$, where $i$ denotes the coordinate of a point and $c$ the color. We transform these sets of points into a one-dimensional point set $S$ in rank space, simply by using the colors as secondary coordinates. More formally, we replace each point $(i, c)$ by the point $(i', c)$ where $i' = |(C_i)_{\leq c}| + \sum_{j=1}^{i-1} |C_j|$. Here $|(C_i)_{\leq c}|$ denotes the number of colors in $C_i$ that are less than or equal to $c$. Finally, we transform $S$ into a two-dimensional point set $\bar{S}$ without colors, by mapping each colored point $(i', c) \in S$ to the two-dimensional point $(i', \text{pred}(i', c))$ where $\text{pred}(i', c)$ denotes the coordinate of the predecessor of $(i', c)$ amongst points in $S$ with color $c$.

To answer a query $(a, b)$, we simply ask the three-sided query $[a', b'] \times (-\infty, a' - 1]$ on $\bar{S}$. Here $a' = 1 + \sum_{j=1}^{a-1} |C_j|$ and $b' = \sum_{j=1}^{b} |C_j|$. The correctness follows from the fact that for each color $c$ with a point in the range $[a, b]$, precisely one point in $S$ with color $c$ is inside the range $[a', b']$, and also has a predecessor of color $c$ before $a'$. It follows that precisely the same point is inside the query range $[a', b'] \times (-\infty, a' - 1]$ in $\bar{S}$, and thus we report one point for each color inside the query range $[a, b]$. If we augment points in $\bar{S}$ with the color of the point mapping to it, this returns the set of colors inside the query range. Note that the values $a'$ and $b'$ can be obtained by table lookups in $O(1)$ I/Os, so we can conclude:

**Theorem 2** *There is a linear space data structure for one-dimensional colored range reporting in rank space that answers queries in $O(1 + k/B)$ I/Os.*

**Colored prefix reporting.** We now consider the following problem: Given a set $S$ of strings, each $O(1)$ disk blocks in length, and a function $c : S \to 2^{\{1, \ldots, \sigma\}}$, support queries of the kind: Given a string $p$, report the set $\bigcup_{x \in S \cap p*} c(x)$, where $p*$ denotes the set of strings with prefix $p$. Building on work of Alstrup et al. [7], Belazzougui et al. [13] have shown the following:

**Theorem 3** *Given a collection $S$ of $n$ strings, there is a linear space data structure that, given a string $p$ of length $O(B)$, returns in $O(1)$ I/Os: 1) The interval of ranks (within $S$) of the strings in $S \cap p*$, and 2) The longest common prefix of the strings in $S \cap p*$.*

In particular, the first item means that we have a reduction of prefix queries to range queries in rank space that works in $O(1)$ I/Os and linear space. Combined with Theorem 2 this implies an optimal solution for colored prefix reporting, for prefixes of length $O(B)$.

## 4 Top-$k$ colored prefix reporting

Suppose that we are interested in reporting just the $k$ first colors in $\{1, \ldots, \sigma\}$ that match a prefix query $p$ (of length $O(B)$), where $k$ is now a *parameter of the data structure*. We use the notation $\text{top}_k(\Sigma)$ to denote the largest $k$ elements of a set $\Sigma$ (where $\text{top}_k(\Sigma) = \Sigma$ if $|\Sigma| < k$). The techniques in the previous sections do not seem to lead to optimal results in the I/O model. However, if we consider a more powerful, yet arguably realistic, model with parallel data access, it turns out that a simple data structure provides optimal bounds.

**Scatter-I/O model.** We consider a special case of the parallel disk model [28] where there are $B$ disks, and each block contains a single word. (Notice that we use $B$ differently than one would for the parallel I/O model.) A single I/O operation thus consists of retrieving or writing $B$ words that may be placed arbitrarily in storage. To distinguish this from a normal I/O operation, we propose the notation *sI/O* (for *scatter* I/Os). This model abstracts (and idealizes) the memory model used by IBM's *Cell* architecture [15], which has been shown to alleviate memory bottlenecks for problems such as BFS [25] that are notoriously hard in the I/O model [22].

## 4.1 Our data structure

We construct a collection $S'_k$ consisting of prefixes of strings in $S$. For each $p \in S'_k$ we store the color set $c_k(p) = \text{top}_k(\bigcup_{x \in S \cap p*} c(x))$ in sorted order. Recall that we use $p*$ to denote the set of all strings with prefix $p$. Given a prefix $p$, there is a minimal subset $S_p \subseteq S'_k \cap p*$ that *covers* $p$ in the sense that any string in $S \cap p*$ has a string in $S_p$ as a prefix. This means that the result of a query for $p$ is $\text{top}_k(\bigcup_{x \in S_p} c(x))$.

**Choice of $S'_k$.** We give an inductive definition of $S'_k$, where prefixes of strings in $S$ are considered in decreasing order. A prefix $p$ is included in $S'_k$ if either $p \in S$ or the following condition holds: $\sum_{x \in S_p} |c_k(x)| > 2|\bigcup_{x \in S_p} c_k(x)|$. Since $S_p$ depends only on prefixes longer than $p$, this is well-defined. Intuitively, we build a data structure for $c_k(p)$ if this will reduce the cost of reading all elements in $c_k(p)$ by a factor of more than 2. This happens if more than half of the elements in the multiset $\bigcup_{x \in S_p} c_k(x)$ are duplicates.

**Space usage.** An accounting argument shows that the total space usage for these lists is $O(\sum_{x \in S} |c_k(x)|)$, i.e., linear in the size of the data: For each $x \in S$ place $|c_k(x)|$ credits on $x$ — this is $\sum_{x \in S} |c_k(x)|$ credits in total. If we build a data structure for a prefix $p$, the lists merged to form it have a total of at least $2|c_k(p)|$ credits. This is enough to pay for the space used, $c_k(p)$ credits, and for placing $|c_k(p)|$ credits on $p$. By induction, we can pay for all lists constructed using the original credits, which implies linear space usage.

**Support structures.** To support efficient search for the nodes that cover a given query we consider the *compacted trie* which consists of all branching nodes in the trie of $S$. Any colored prefix query $p_{\text{orig}}$ can be converted into a query for $p$, where $p$ is the path leading to the highest branching node having $p_{\text{orig}}$ as a prefix. By Theorem 3 we can find the string $p$ in $O(1)$ I/Os, using that $p$ has length $O(B)$. For each branching node $p$ we store the sequence of pointers to the nodes $S_p \subseteq S'_k$ that cover $p$. There can be at most $O(B)$ pointers to a given node, one for each ancestor in the trie, i.e., $O(mB)$ pointers in total. Also, we store the number of elements that should be reported from each node in $S_p$. More specifically, we store the list of *non-zero* numbers, with corresponding pointers. The total space for this is $O(mB)$.

**Queries.** On a query we first locate an equivalent branching prefix $p$, as outlined above. Then we retrieve the list of nodes in $S_p$ from which results should be retrieved, and the number of elements from each. This requires $O(1 + k/B)$ I/Os. Note that the total number of elements, including duplicates, is at most $2k$, because if it was larger a merged list would have been created. Finally, since we now know $O(k)$ positions in memory containing all $k$ elements to be reported, it is trivial to retrieve them in $O(k/B)$ sI/Os. This is the only point where we use the extra power of the model — all previous steps involve standard I/Os.

## 5 Open problems

An obvious question is if our results for the I/O model can be extended to the cache-oblivious model [18]. Also, it would be interesting to investigate whether our results can be obtained with the indivisibility assumption, or if the problem separates the I/O model with and without the indivisibility assumption. Finally, it would be interesting to see if top-$k$ colored prefix (and range) reporting admits an efficient solution in the I/O model, or if the top-$k$ version separates the I/O and scatter I/O models.

**Acknowledgement.** We thank for ChengXiang Zhai for making us aware of query relaxation in information retrieval.